%% file: main.tex
\documentclass[conference]{IEEEtran}
\IEEEoverridecommandlockouts

\usepackage{cite}
\usepackage{amsmath,amssymb,amsfonts}
\usepackage{algorithmic}
\usepackage{graphicx}
\usepackage{textcomp}
\usepackage{xcolor}
\usepackage{hyperref}
\usepackage{booktabs}
\usepackage{subcaption}
\usepackage{xurl}  
\usepackage{multirow}
\usepackage{tikz}
\usetikzlibrary{positioning, calc, arrows.meta}
\usepackage{pgfplots}
\pgfplotsset{compat=1.18}
\usepackage{balance}

\begin{document}

\title{Bounded-Memory Parallel Image Pulling for Large Container Images}

\author{
\IEEEauthorblockN{Sri Saran Balaji Vellore Rajakumar, \quad Henry Wang, \quad Ankur Singh, \quad James Thompson}
\IEEEauthorblockA{\textit{Amazon Web Services} \\
Seattle, USA}
}

\maketitle

\begin{abstract}
AI/ML workloads increasingly run as containers, where a container image must be downloaded to the host before the workload can start. This cold image pull lands on the critical path whenever a training or inference job scales up or a host is updated, and for GPU workloads it has become the dominant component of startup time as AI/ML images reach 31--48~GiB compressed. We present Disk-Backed Parallel Pull (DBPP), an alternative to the in-memory ordered reassembly used by containerd~2.2, the upstream container runtime. containerd splits layers into chunks fetched concurrently over HTTP range requests, but chunks that arrive out of order accumulate in the runtime heap until a sequential consumer drains them in order. This backlog grows with image size, and on GPU nodes where host memory is shared with frameworks and model weights, it leads to out of memory (OOM) termination of the runtime itself.

DBPP writes each chunk directly to its target byte offset on disk, eliminating the ordering dependency and bounding memory regardless of image size. Because each layer lands on disk as a complete, seekable file, DBPP runs SHA-256 digest verification and decompression simultaneously, two passes containerd must run one after the other. In controlled experiments across five production-scale images (up to 48.5~GiB), DBPP reduces peak daemon memory by 8.7--25.3$\times$ while maintaining comparable pull throughput. On a memory-constrained node, containerd~2.2 is OOM-killed pulling a 31.4~GiB image while DBPP completes the same pull. The underlying idea reaches past container images: any pipeline that buffers data in memory only to enforce ordering can move that buffer to disk once the backing store is fast enough, trading a scarce, contended resource for an abundant one.
\end{abstract}


\input{sections/introduction}
\input{sections/characterization}
\input{sections/related_work}
\input{sections/design}
\input{sections/evaluation}
\input{sections/discussion}
\input{sections/conclusion}

\balance
\bibliographystyle{IEEEtran}
\bibliography{references}

\end{document}

%% file: sections/introduction.tex
\section{Introduction}
\label{sec:introduction}

Large-scale AI and ML training and inference workloads increasingly ship as container images. Frameworks such as PyTorch and Megatron bundle GPU runtimes, numerical libraries, and model-serving code into images that reach 31--48~GiB compressed (Table~\ref{tab:image-survey}). These images are deployed on GPU-accelerated cloud instances with 100--3200~Gbps of network bandwidth, enough to transfer them in seconds. But in practice, the pull takes far longer, because the container runtime must also reassemble and unpack each layer before the image is ready to run. The bottleneck is not the network but how the runtime assembles the image.

When a container image is pulled, the runtime fetches each layer from a remote registry over HTTP, verifies the layer against its cryptographic digest, decompresses it, and unpacks the result into a local filesystem. For large images this process dominates pod startup, accounting for more than 75\% of total startup time~\cite{awssocipull2025}. The pull path is the critical path.

The containerd container runtime~\cite{containerd2024} addressed part of this by introducing intra-layer parallelism: splitting each layer into fixed-size chunks and fetching them concurrently via HTTP range requests. However, its reassembly buffers chunks in an ordered, in-memory pipe: chunks that arrive early are held in the daemon heap until a sequential consumer reaches their position. Because parallel download outruns that consumer, the backlog grows with image size. On memory-constrained GPU nodes (where system memory is shared with CUDA contexts and model weights), this exhausts the host: pulling a 31.4~GiB image on a node with 7.6~GiB random-access memory (RAM) terminates the container runtime daemon via the kernel OOM killer.

We present Disk-Backed Parallel Pull (DBPP), which we have implemented in the Seekable Open Container Initiative (SOCI) snapshotter~\cite{soci2023}. Instead of buffering chunks in memory, each chunk is written directly to its byte offset on disk via kernel system calls. The ordering dependency is removed and daemon memory stays bounded regardless of image size. The design operates on unmodified OCI images~\cite{ociimage2023} with standard HTTP range requests~\cite{ocidist2023}, requiring no registry-side changes.

Section~\ref{sec:characterization} characterizes the structural properties of production ML images. Section~\ref{sec:related} surveys existing approaches to container image distribution. Section~\ref{sec:design} presents the DBPP design and its two-digest integrity scheme. Section~\ref{sec:evaluation} evaluates DBPP against containerd~2.2 (in-memory ordered reassembly) across a corpus of five production-scale images, measuring memory consumption, pull time, and behavior under memory pressure.

%% file: sections/characterization.tex
\section{Background}
\label{sec:characterization}

An image pull is a few steps in a pipeline. For large ML images, one of them, stitching a layer back together from concurrently fetched chunks, dominates everything else. This section sets up why.

\subsection{How a Container Image Pull Works}
\label{sec:pull-primer}

A container image is a bundle of read-only filesystem \emph{layers} plus a \emph{manifest} that lists them. Each layer is a compressed (gzip) tar archive, and each is named by the cryptographic hash (SHA-256) of its bytes, called its \emph{digest}. The digest names the layer, so hosts can cache and deduplicate it, and it lets a host verify the bytes it received are exactly the bytes the author published.

Images are stored in a \emph{registry}, an HTTP service such as Amazon Elastic Container Registry (ECR) or Docker Hub. Before a container can start, the host must \emph{pull} the image. For each layer it downloads the compressed bytes from the registry over HTTP, verifies them against the layer's digest, decompresses the archive, and unpacks the files into a local directory that becomes part of the container's root filesystem. The container cannot start until every layer has been pulled, verified, and unpacked. For the large images we study, this pull phase dominates startup time.

On most hosts this pipeline is driven by \emph{containerd}, a container runtime daemon used by Kubernetes, Docker, and Amazon Elastic Container Service (ECS). containerd delegates the on-disk management of layers to a pluggable component called a \emph{snapshotter}. This paper is implemented in the SOCI (Seekable OCI) snapshotter, a plugin that layers on top of containerd without modifying it.

A naive pull restricts throughput to the maximum capacity of a single TCP stream, leaving the 100 Gbps network link severely underutilized. The remedy is \emph{intra-layer parallelism}: a large layer is split into fixed-size \emph{chunks}, and the chunks are fetched concurrently using HTTP \emph{range requests} (a standard mechanism for asking a server for a specific byte range of a file). Those chunks land out of order, so something must \emph{reassemble} them into the original layer before verification and decompression. How that reassembly is done, in memory or on disk, is the subject of this paper.

\subsection{ML Container Images Are Large and Skewed}
\label{sec:image-survey}

Production ML images have structural properties that stress the reassembly step. Table~\ref{tab:image-survey} reports sizes measured directly from registry manifests for the \texttt{linux/amd64} platform; all are publicly available and independently reproducible.

\begin{table}[t]
    \centering
    \scriptsize
    \caption{Production ML images, measured from registry manifests (compressed, linux/amd64). Sizes in binary GiB.}
    \label{tab:image-survey}
    \begin{tabular}{@{}lrr@{}}
        \toprule
        \textbf{Image (registry / tag)} & \textbf{GiB} & \textbf{Lyr} \\
        \midrule
        rocm/pytorch-training:v26.1~\cite{rocm2024} & 48.5 & 99 \\
        rocm/megatron-lm:v26.1 & 48.5 & 99 \\
        nvcr.io/nvidia/nemo:24.09 & 31.4 & 111 \\
        rocm/pytorch-training:v25.5 & 30.5 & 80 \\
        rocm/megatron-lm:v25.8\_py310 & 29.0 & 99 \\
        AWS SGLang DLC & 19.3 & 61 \\
        pytorch/pytorch:2.11.0-cuda12.8-cudnn9-devel & 13.1 & 10 \\
        AWS vLLM DLC & 9.7 & 38 \\
        AWS PyTorch-training DLC 2.4.0-gpu & 9.1 & 28 \\
        \bottomrule
    \end{tabular}
\end{table}

Tens of gibibytes is the norm; multiple default tags exceed 28~GiB compressed. The distribution is bimodal: a handful of multi-gigabyte layers dominate an image built from dozens of small ones (NeMo packs 111 layers, several over 2~GiB). CUDA and Python binaries expand about threefold on decompression while already-compressed model weights barely grow, so the balance between download cost and unpack cost shifts from one image to the next.

That bimodality dictates the download strategy. Fetching many layers at once (inter-layer parallelism) helps only up to the size of the largest single layer: a 2~GiB layer on one connection stalls the whole pull. Saturating a fast network therefore means splitting each large layer into chunks and fetching them concurrently, the intra-layer parallelism from above. Both containerd and our design do this. They differ in one thing: how those chunks get reassembled. That one choice decides the memory behavior.

%% file: sections/related_work.tex
\section{Related Work}
\label{sec:related}

Prior work on container startup acceleration differs along four axes: whether it requires a modified image format, whether it requires a custom registry, whether it improves a single cold node or the whole cluster, and whether it targets dense or sparse image access. Table~\ref{tab:related} sorts the main families along these axes. Only one prior system is directly comparable to DBPP: containerd's in-memory parallel pull. It fetches the same unmodified OCI images from the same registries over the same protocol, and differs from DBPP in exactly one thing: how chunks get reassembled. Comparing DBPP against a format-changing system would not isolate the reassembly strategy, so the evaluation measures against containerd and places all other approaches qualitatively.

\begin{table}[t]
    \centering
    \scriptsize
    \caption{Where DBPP sits among startup-acceleration approaches.}
    \label{tab:related}
    \begin{tabular}{@{}lccccc@{}}
        \toprule
        \textbf{Approach} & \textbf{Std.} & \textbf{Std.} & \textbf{Cold} & \textbf{Dense} & \textbf{Bounded} \\
        & \textbf{format} & \textbf{registry} & \textbf{node} & \textbf{access} & \textbf{memory} \\
        \midrule
        Lazy loading (SOCI, eStargz) & \checkmark & \checkmark & \checkmark & --- & \checkmark \\
        Format redesign (DADI, Nydus) & --- & --- & \checkmark & \checkmark & \checkmark \\
        P2P (Dragonfly, Kraken) & \checkmark & --- & --- & \checkmark & --- \\
        Pre-warming (Bottlerocket) & \checkmark & \checkmark & --- & \checkmark & \checkmark \\
        containerd in-memory pull & \checkmark & \checkmark & \checkmark & \checkmark & --- \\
        \textbf{DBPP (this work)} & \checkmark & \checkmark & \checkmark & \checkmark & \checkmark \\
        \bottomrule
    \end{tabular}
\end{table}

Slacker~\cite{harter2016slacker} found that only 6.4\% of image data is accessed during execution, motivating \emph{lazy loading}. SOCI~\cite{soci2023}, eStargz~\cite{estargz2020}, and AWS Lambda's loader~\cite{brooker2023lambda} bring lazy loading to OCI images. These are complementary to DBPP. DBPP handles the dense case, where most bytes are actually needed. Both modes live inside the same SOCI snapshotter and are controlled by a single configuration flag, so switching between lazy loading and full parallel pull requires no image rebuilds, registry changes, or runtime upgrades.

DADI~\cite{li2020dadi}, Nydus~\cite{nydus2023}, and FlacIO~\cite{flacio2025} redesign the image format itself for parallel or block-level access. But they need format changes or a custom registry; DBPP runs on unmodified OCI images with plain HTTP range requests.

Dragonfly~\cite{dragonfly2023}, Kraken~\cite{kraken2019}, FaaSNet~\cite{faasnet2021}, and Starlight~\cite{chen2022starlight} distribute images peer-to-peer to optimize aggregate cluster bandwidth. On a cold autoscale event, no peer has the image yet, so they do nothing for a single node's speed. DBPP's per-node pull optimization is orthogonal and stacks on top of P2P distribution.

Default pull clients such as Docker~\cite{moby2024}, containerd~\cite{containerd2024}, and skopeo~\cite{containersimage2024} download multiple layers simultaneously but fetch each individual layer as a single sequential stream over one connection. The only remedy is to split the large layer itself into fixed-size chunks and fetch all chunks concurrently over separate connections, which is \emph{intra}-layer parallelism. Among the approaches surveyed here, containerd~2.2 and DBPP are the only ones that do this without requiring image format changes or a custom registry. That is why containerd~2.2 is the head-to-head baseline: it attacks the same layer the same way, and differs from DBPP only in where the chunks land.

Bottlerocket~\cite{aws2024bottlerocket} and DaemonSet pre-pullers eliminate pull time for predictable workloads, but provide no benefit for a new image version or an unpredicted spot replacement: exactly the cold-pull cases DBPP targets.

DBPP is deployed in production at scale. Amazon Elastic Kubernetes Service (Amazon EKS) Auto Mode enables DBPP by default on GPU instances~\cite{eksautomode2025,eksultrascale2025}, and outside adopters report large startup gains from switching to parallel pull: AppsFlyer an 82\% cut in cold-start time~\cite{appsflyer2025soci} and Grab a 60\% cut in image download time~\cite{grab2025lazyloading}. Against containerd~2.2 as a parallel-pull baseline, DBPP holds pull time within 12\% (Section~\ref{sec:eval-pulltime}) and its win is bounded memory, not additional speed.

%% file: sections/design.tex
\section{Design}
\label{sec:design}

We implement disk-backed parallel pull, which we call DBPP, as a mode of the SOCI snapshotter~\cite{soci2023}, a containerd plugin. An operator can enable DBPP on a node without patching or upgrading containerd, because SOCI is a containerd plugin that extends the runtime without modifying it. DBPP operates on standard OCI images~\cite{ociimage2023} fetched with ordinary HTTP range requests and requires no changes on the registry side. It ships upstream and is enabled by default on Amazon EKS optimized machine images for GPU instances.

Two designs reassemble concurrently fetched chunks differently: containerd's ordered in-memory approach, and the disk-backed approach DBPP introduces. The full source is in the SOCI repository~\cite{soci2023}; we describe just enough of the mechanism to explain why memory behaves the way it does.

\subsection{Two Ways to Reassemble a Layer}
\label{sec:architecture}

\begin{figure*}[t]
    \centering
    \begin{tikzpicture}[
        node distance=0.6cm and 2.5cm,
        box/.style={draw, rounded corners, minimum width=2.2cm, minimum height=0.6cm, font=\scriptsize, align=center},
        arr/.style={->, >=stealth, thick},
        label/.style={font=\scriptsize\itshape, text=black!50},
        title/.style={font=\small\bfseries}
    ]
    \node[title] (titleL) {In-memory ordered reassembly};
    \node[box, below=0.4cm of titleL, fill=black!5] (regL) {Registry};
    \node[box, below=of regL, fill=black!5] (chunksL) {N concurrent\\range requests};
    \node[box, below=of chunksL, fill=black!20, thick] (pipesL) {Per-chunk buffers\\held in memory};
    \node[box, below=of pipesL, fill=black!20, thick] (readerL) {Single consumer\\reads strictly in order};
    \node[box, below=of readerL, fill=black!8] (storeL) {Write to store\\+ hash inline};
    \node[box, below=of storeL, fill=black!8] (readbackL) {Read layer back from disk};
    \node[box, below=of readbackL, fill=black!8] (unpackL) {Decompress + extract\\+ verify uncompressed digest};

    \draw[arr] (regL) -- (chunksL);
    \draw[arr] (chunksL) -- node[right, label] {chunks buffer} (pipesL);
    \draw[arr] (pipesL) -- node[right, label] {drain in order} (readerL);
    \draw[arr] (readerL) -- (storeL);
    \draw[arr] (storeL) -- node[right, label] {serial} (readbackL);
    \draw[arr] (readbackL) -- (unpackL);

    \node[right=0.3cm of pipesL, font=\scriptsize\bfseries, align=left] {$\leftarrow$ early chunks\\[-2pt]\hspace{4pt}wait here};

    \node[title, right=5.5cm of titleL] (titleR) {Disk-backed parallel pull (DBPP)};
    \node[box, below=0.4cm of titleR, fill=black!5] (regR) {Registry};
    \node[box, below=of regR, fill=black!5] (chunksR) {N concurrent\\range requests};
    \node[box, below=of chunksR, fill=black!8, dashed] (pwriteR) {Each chunk written to\\its byte offset (any order)};
    \node[box, below=1.0cm of pwriteR, fill=black!8] (fileR) {Complete layer file on disk};
    \node[box, below left=0.8cm and 0.3cm of fileR, fill=black!5] (hashR) {Read file $\rightarrow$\\verify compressed digest};
    \node[box, below right=0.8cm and 0.3cm of fileR, fill=black!5] (unpackR) {Read file $\rightarrow$ decompress\\+ verify uncompressed digest};

    \draw[arr] (regR) -- (chunksR);
    \draw[arr] (chunksR) -- node[right, label] {no buffering} (pwriteR);
    \draw[arr] (pwriteR) -- node[right, label] {no ordering} (fileR);
    \draw[arr] (fileR) -- (hashR);
    \draw[arr] (fileR) -- (unpackR);

    \node[below=2.4cm of fileR, font=\scriptsize\bfseries] (concLabel) {concurrent};
    \draw[dotted, thick] (hashR.south) -- ++(0,-0.15) -| (concLabel.west);
    \draw[dotted, thick] (unpackR.south) -- ++(0,-0.15) -| (concLabel.east);

    \node[right=0.3cm of pwriteR, font=\scriptsize\itshape, align=left] {$\leftarrow$ never\\[-2pt]\hspace{4pt}waits};

    \node[font=\large, text=black!40] at ($(pipesL.east)!0.5!(pwriteR.west)+(0,0.3)$) {vs.};
    \end{tikzpicture}
    \caption{In-memory ordered reassembly (left) vs.\ disk-backed parallel pull (right). Both fetch chunks concurrently; they diverge in how chunks are reassembled and how verification and decompression are scheduled.}
    \label{fig:architecture}
\end{figure*}
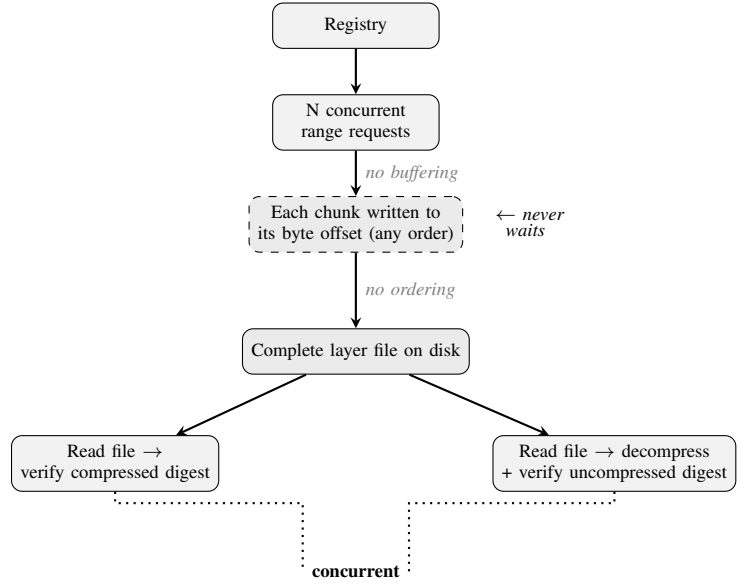

Both designs split a layer into chunks of a fixed size and fetch many chunks at once over separate HTTP range requests (Figure~\ref{fig:architecture}). They differ entirely in what happens as those chunks arrive.

\paragraph{In-memory ordered reassembly}
containerd (in-memory) keeps each arriving chunk in a memory buffer and feeds all chunks to a single consumer that must read them strictly in order: it consumes chunk~0 to its end, then chunk~1, and so on. Reading in order is what lets the consumer compute the layer's SHA-256 digest as a running hash while it writes the bytes into containerd's content store. The catch is that a chunk which finishes downloading early cannot be handed to the consumer until every chunk before it has been consumed, so it sits in memory and waits. Downloads run in parallel, but one consumer drains them one at a time. So downloads outrun the consumer, and finished chunks pile up waiting. This waiting set is a live backlog, not a leak: it persists for the whole pull, and its size tracks the image. The compressed-layer SHA-256 is computed for free as part of this ordered stream. After the consumer finishes, a single read-back pass decompresses the stored layer, extracts the filesystem, and computes the DiffID over the decompressed bytes.

\paragraph{Disk-backed parallel pull}
DBPP removes the ordering constraint. The destination file is pre-sized to the layer's full length so that every byte position exists before any download starts. Each chunk gets written straight to its own offset the moment it arrives, with \texttt{pwrite}. Chunk~37 lands at its offset whether or not chunks~0 through~36 have arrived, and its bytes leave process memory the moment they are written. No chunk waits for another, and nothing accumulates in the heap. Local NVMe absorbs writes at $\sim$3~GiB/s, far faster than any single connection delivers its bytes, so the disk is never the bottleneck.

\subsection{Preserving Per-Layer Integrity}
\label{sec:verification}

OCI requires each layer to be verified against two digests: one over the compressed bytes (the layer's registry identity) and one over the decompressed bytes (its \emph{DiffID}, which ties the extracted filesystem to the image configuration). containerd (in-memory) computes the compressed digest for free as a running hash during the ordered download stream. The DiffID requires a separate pass: the layer must be fully written to the content store first, then a single read-back decompresses it and computes the DiffID over the output.

Out-of-order arrival means DBPP cannot compute a running hash during download, so it must read the assembled layer back from disk to verify it. But because the layer is now a complete, seekable file, two readers can sweep it at once: one re-reads the compressed bytes to verify the layer digest, while the other decompresses the file and verifies the DiffID as the decompressed bytes stream to the unpacker. containerd (in-memory) gets the compressed digest for free during download and pays only one read-back for decompression and DiffID. DBPP must do an extra read of the compressed layer for the SHA-256 that containerd computes inline, but it recovers that cost by running the compressed-digest read and the decompression read concurrently rather than sequentially, so it spends more instantaneous CPU but adds no wall-clock time (Section~\ref{sec:eval-cpu}).

\subsection{Parallel Unpack and Scope}
\label{sec:parallel-unpack}

Because each layer becomes an independent directory under the overlayfs snapshotter, multiple layers can decompress and unpack at the same time, bounded by a configurable concurrency limit. We enable this parallel unpack on both designs so that it is not a confounding variable in our comparison. Decompression \emph{within} a single layer, however, remains serial: \texttt{unpigz} degrades to single-core inflation for a single large layer because container layers are monolithic gzip streams with no independent members to parallelize across. Once download is parallel, that serial per-layer decompression is what dominates; we return to it in Section~\ref{sec:discussion}.

%% file: sections/evaluation.tex
\section{Evaluation}
\label{sec:evaluation}

Disk-backed parallel pull takes the same wall-clock time as containerd~2.2 (in-memory) across the full corpus and uses comparable CPU on real images. In exchange it cuts daemon memory from 4.1--19.2~GiB down to 243~MiB--1.02~GiB, flat regardless of image size. On a node where the baseline OOM-kills, DBPP finishes the pull.

\subsection{Methodology}
\label{sec:eval-methodology}

We change only how chunks are reassembled and hold everything else fixed, so any measured difference is attributable to that one choice. Both designs run parallel chunked download and parallel unpack with the same chunk size, download and unpack concurrency, decompressor, instance type, registry, region, and pull client. Table~\ref{tab:controlled} lists the fixed parameters. The two stacks run on separate but identically configured instances in the same availability zone. We use a large-memory instance (256~GiB) so containerd (in-memory) can \emph{complete} and we can measure its actual peak, not just watch it die. The small-node experiment is Section~\ref{sec:eval-constrained}.

\begin{table}[t]
    \centering
    \footnotesize
    \caption{Parameters held identical across both designs.}
    \label{tab:controlled}
    \begin{tabular}{@{}ll@{}}
        \toprule
        \textbf{Parameter} & \textbf{Value (both designs)} \\
        \midrule
        Chunk size & 16~MiB \\
        Concurrent downloads & 40 \\
        Concurrent unpacks & 50 \\
        Decompressor & \texttt{unpigz} (single-core per layer; see \S\ref{sec:discussion}) \\
        Instance type & m6idn.16xlarge (64 vCPU, 256 GiB) \\
        Network & up to 100~Gbps \\
        Storage & Local NVMe (both roots, same device) \\
        Registry & ECR, same region (us-west-2) \\
        Pull client & \texttt{crictl} (the path the kubelet uses) \\
        containerd / SOCI & 2.2.4 / 0.14.0 \\
        \bottomrule
    \end{tabular}
\end{table}

Every pull is a true cold pull: we stop the daemons, wipe all storage roots, drop the kernel page cache, and restart before each run. We use \texttt{crictl} because it is the only client that drives both parallel download and parallel unpack on containerd~2.2. We report three metrics. \emph{Pull time} is wall-clock from pull start to completion. \emph{$\Delta$RSS} is the peak growth in resident memory of the pull-path daemons over their idle baseline, sampled at 5~Hz. \emph{CPU-seconds} is total non-idle CPU time over the pull. All sizes are in binary GiB (1~GiB~=~$2^{30}$~bytes).

Table~\ref{tab:corpus} lists the corpus: a two-layer incompressible synthetic (one 2.2~MiB busybox base plus one 5.00~GiB payload) to isolate raw download bandwidth, a nine-layer synthetic (one 2.2~MiB busybox base plus eight large payload layers, 50.0~GiB total) to stress both intra- and inter-layer parallelism at scale, and three production ML images spanning 7.1 to 48.5~GiB.

\begin{table}[t]
    \centering
    \scriptsize
    \caption{Evaluation corpus. Sizes are compressed, in binary GiB.}
    \label{tab:corpus}
    \begin{tabular}{@{}lrrp{2.4cm}@{}}
        \toprule
        \textbf{Image} & \textbf{Size (GiB)} & \textbf{Lyr} & \textbf{Purpose} \\
        \midrule
        Synthetic, incompressible & 5.00 & 2 & Isolate intra-layer download \\
        SageMaker Distrib.\ 3.1-gpu & 7.06 & 24 & Real, one dominant layer \\
        NVIDIA NeMo 24.09 & 31.4 & 111 & Real, many-layer + large \\
        ROCm PyTorch-training v26.1 & 48.5 & 99 & Real, largest in survey \\
        Synthetic, multi-large-layer & 50.0 & 9 & Stress inter+intra at scale \\
        \bottomrule
    \end{tabular}
\end{table}

\subsection{Pull Time Is Within Noise}
\label{sec:eval-pulltime}

Pull times are within 12\% of each other across the whole corpus in both directions (Table~\ref{tab:pulltime}). DBPP is faster on the three smaller images and slower on the two largest; neither gap is consistent or large enough to be meaningful. The buffering strategy does not touch the bottleneck: on real images the pull is dominated by serial single-core decompression, which both designs hand to the same decompressor, so where the chunks are buffered cannot change the total.

\begin{table}[t]
    \centering
    \caption{Pull time, median [min, max] over $n{=}5$ cold runs per stack. Ratio is ctrd/DBPP median.}
    \label{tab:pulltime}
    \scriptsize
    \begin{tabular}{@{}lrrr@{}}
        \toprule
        \textbf{Image} & \textbf{ctrd (s)} & \textbf{DBPP (s)} & \textbf{Ratio} \\
        \midrule
        Synthetic (5.00 GiB)   & 17.3 [17.0, 17.9]    & 15.5 [15.4, 15.6]    & 1.12$\times$ \\
        SageMaker (7.06 GiB)   & 113.5 [112.3, 114.0] & 104.5 [103.7, 109.6] & 1.09$\times$ \\
        NeMo (31.4 GiB)        & 116.6 [113.3, 119.8] & 107.1 [106.4, 107.7] & 1.09$\times$ \\
        ROCm (48.5 GiB)        & 157.7 [151.8, 160.6] & 160.5 [155.9, 162.9] & 0.98$\times$ \\
        Synth-multi (50.0 GiB) & 63.9 [62.8, 65.1]    & 69.4 [68.8, 71.5]    & 0.92$\times$ \\
        \bottomrule
    \end{tabular}
\end{table}

\subsection{Memory Stays Flat as Images Grow}
\label{sec:eval-memory}

containerd's resident set (in-memory) grows with the image: 14.7~GiB on the 31~GiB NeMo, and 19.2~GiB on the 48.5~GiB ROCm image. The cause is the backlog. Parallel chunk downloads finish faster than the single ordered consumer can drain them, so fully-downloaded chunks sit in heap waiting their turn. The pile is proportional to image size because a larger image means more chunks ahead of the consumer at any moment.

\begin{figure}[t]
\centering
\includegraphics[width=\columnwidth]{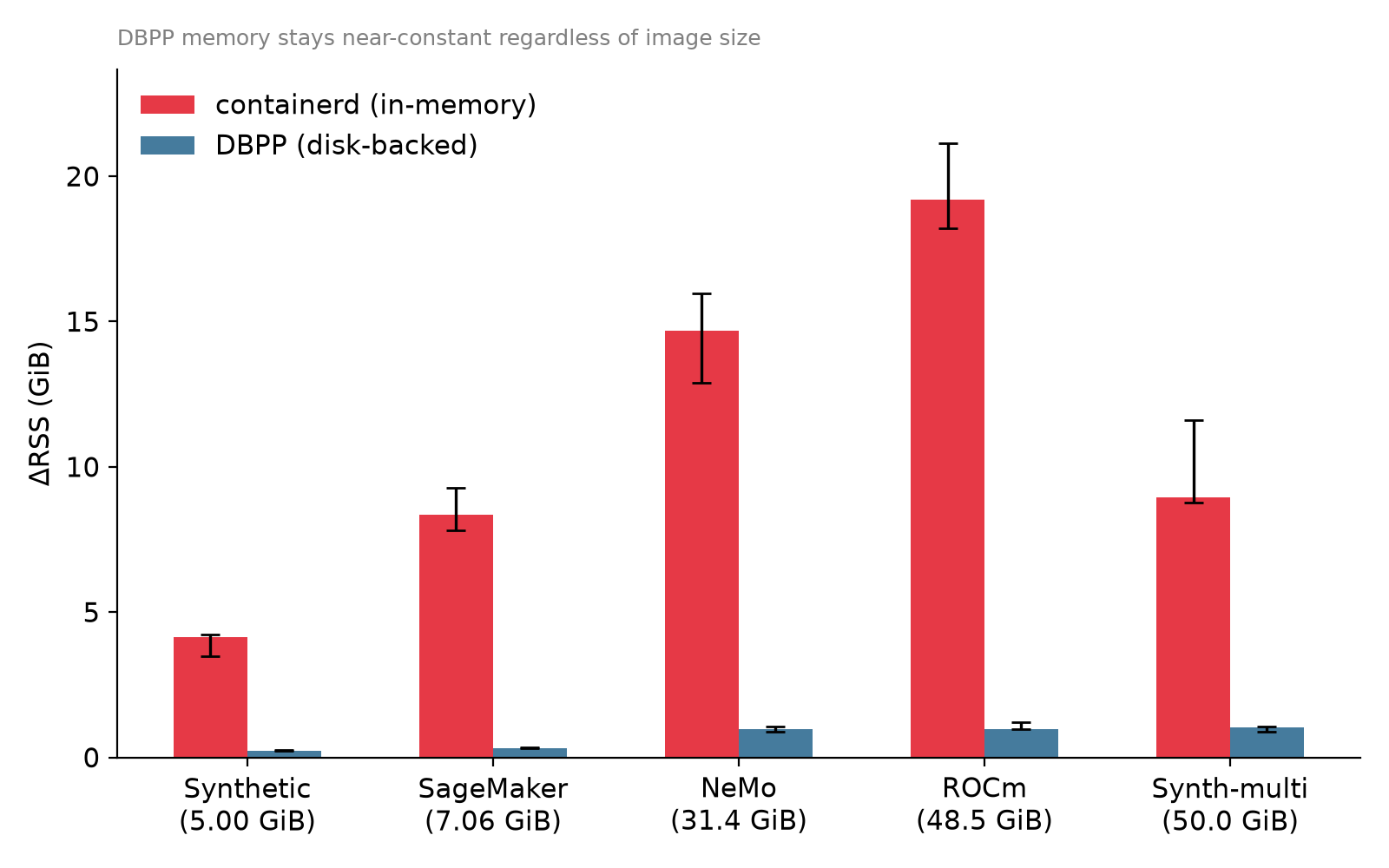}
\caption{Peak daemon $\Delta$RSS during pull. containerd (in-memory) scales with image size (4.1--19.2~GiB); DBPP stays near-constant (243~MiB--1.02~GiB).}
\label{fig:memory}
\end{figure}

DBPP eliminates the backlog at the source. Each chunk calls \texttt{pwrite} to its fixed byte offset and returns; the bytes leave process memory immediately. There is no ordering dependency and no waiting pile, so daemon memory stays between 243~MiB and 1.02~GiB across the entire corpus regardless of image size (Figure~\ref{fig:memory}). The reduction is 8.7$\times$ to 25.3$\times$ depending on the image, but the absolute number is what matters: around a gigabyte no matter how large the image.

The per-process breakdown makes the mechanism visible directly in the data (Table~\ref{tab:memory-results}). On the baseline, all of the growth is in containerd, the single binary that buffers, orders, hashes, and unpacks. On DBPP, containerd stays between 126~MiB and 381~MiB across all five images, holding only gRPC coordination and snapshot metadata. The soci-snapshotter process carries 119~MiB to 828~MiB of streaming I/O buffers and goroutine state. The backlog does not move to the snapshotter; it disappears.

\begin{table}[t]
    \centering
    \scriptsize
    \caption{Peak $\Delta$RSS by process, median of $n{=}5$; run-to-run spread is shown in Figure~\ref{fig:memory}. On DBPP, containerd is a bystander; growth is in the snapshotter's I/O buffers. Per-process peaks are not simultaneous; Ratio is baseline ctrd / peak combined DBPP RSS.}
    \label{tab:memory-results}
    \begin{tabular}{@{}lrrrr@{}}
        \toprule
        \textbf{Image} & \textbf{ctrd (bl)} & \textbf{ctrd (DBPP)} & \textbf{soci (DBPP)} & \textbf{Ratio} \\
        \midrule
        Synthetic (5.00 GiB)   & 4{,}241~MiB & 126~MiB & 119~MiB & 17.5$\times$ \\
        SageMaker (7.06 GiB)   & 8{,}553~MiB & 157~MiB & 222~MiB & 25.3$\times$ \\
        NeMo (31.4 GiB)        & 15{,}040~MiB & 362~MiB & 744~MiB & 15.0$\times$ \\
        ROCm (48.5 GiB)        & 19{,}647~MiB & 381~MiB & 828~MiB & 19.4$\times$ \\
        Synth-multi (50.0 GiB) & 9{,}162~MiB & 361~MiB & 702~MiB & 8.7$\times$ \\
        \bottomrule
    \end{tabular}
\end{table}

\subsection{On a Memory-Constrained Node, the Baseline Fails}
\label{sec:eval-constrained}

The numbers above were collected on a 256~GiB instance specifically so the baseline could complete. On the kind of node these images actually run on, it cannot. We pull the 31.4~GiB NeMo image on an instance with 7.6~GiB of usable RAM and no swap. The baseline's resident set outgrows available memory and the kernel OOM-killer fires on all three trials:

{\footnotesize
\begin{verbatim}
Out of memory: Killed process 26119 (containerd)
total-vm:10290072kB, anon-rss:7516860kB,
... oom_score_adj:-999
\end{verbatim}
}

\noindent The kernel reports \texttt{anon-rss} in KiB despite the \texttt{kB} label, so containerd died holding $7{,}516{,}860 / 1024 = 7{,}341$~MiB ($7.17$~GiB) of anonymous, non-reclaimable heap: the chunk backlog. It held the OOM-kill protection score of $-999$, reserved for the most critical processes, and the kernel killed it anyway after exhausting every lower-priority process. The pull never completed. DBPP completed all three trials, peaking at 3.5~GiB of daemon $\Delta$RSS (Table~\ref{tab:constrained}). DBPP's footprint here is larger than the $\sim$1~GiB it uses on the 64-core node (Table~\ref{tab:memory-results}) because this instance has only 2~vCPUs: single-core decompression runs $5\times$ longer (540~s vs 106~s), so the snapshotter's transient working set is live far longer and accumulates a higher peak. It is still less than half of host memory, which is why the pull completes; the backlog the baseline cannot avoid is exactly the memory DBPP never allocates.

\begin{table}[t]
    \centering
    \footnotesize
    \caption{NeMo 31.4~GiB pull on a memory-constrained node (m6id.large, 7.6~GiB usable RAM, no swap). Three trials each.}
    \label{tab:constrained}
    \begin{tabular}{@{}lrr@{}}
        \toprule
        \textbf{Metric} & \textbf{containerd} & \textbf{DBPP} \\
        \midrule
        Outcome              & OOM-killed (3/3) & completed (3/3) \\
        containerd RSS       & 7{,}341~MiB (at death) & 133~MiB (at end) \\
        Peak daemon $\Delta$RSS & --- & 3{,}589~MiB \\
        Pull completed?      & No (died at 27~s) & Yes (540~s) \\
        \bottomrule
    \end{tabular}
\end{table}

\subsection{CPU Is Comparable on Real Images}
\label{sec:eval-cpu}

CPU usage tells the same story as pull time: on real ML images the two designs are within noise (Table~\ref{tab:cpu-results}). DBPP does an extra sequential read of each compressed layer for digest verification, a pass containerd folds into its ordered stream for free. On the incompressible synthetic there is almost no decompression work, so this extra read runs on its own and shows up directly as added CPU: a 2.9$\times$ ratio. On the real images decompression is the dominant cost and runs concurrently with the verification read, so the extra read overlaps it and the two stacks converge: NeMo 0.98$\times$, ROCm 0.95$\times$. Both leave the 64-core machine roughly 90\% idle throughout, so the verification pass is spending cores that would otherwise sit unused.

\begin{table}[t]
    \centering
    \caption{CPU-seconds during pull (total non-idle CPU time), median [min, max] over $n{=}5$. Ratio is DBPP/ctrd median.}
    \label{tab:cpu-results}
    \scriptsize
    \begin{tabular}{@{}lrrr@{}}
        \toprule
        \textbf{Image} & \textbf{ctrd} & \textbf{DBPP} & \textbf{Ratio} \\
        \midrule
        Synthetic (5.00 GiB)    & 39.5 [38.2, 40.4]      & 113.0 [95.0, 121.7]    & 2.9$\times$ \\
        SageMaker (7.06 GiB)    & 241.9 [239.4, 248.0]   & 325.0 [300.3, 361.1]   & 1.3$\times$ \\
        NeMo (31.4 GiB)         & 973.9 [947.3, 995.6]   & 958.5 [941.5, 963.2]   & 0.98$\times$ \\
        ROCm (48.5 GiB)         & 1{,}401 [1{,}396, 1{,}421] & 1{,}336 [1{,}304, 1{,}351] & 0.95$\times$ \\
        Synth-multi (50.0 GiB)  & 384.6 [380.8, 387.8]   & 543.8 [539.6, 558.6]   & 1.4$\times$ \\
        \bottomrule
    \end{tabular}
\end{table}

%% file: sections/discussion.tex
\section{Discussion and Limitations}
\label{sec:discussion}

\subsection{Disk as a Buffer for Memory-Bound Pipelines}

This pattern generalizes beyond container images. A pipeline that buffers data in memory only to satisfy an ordering or reassembly constraint pays a memory cost that scales with its input. When the backing store is fast enough that writing and re-reading stays off the critical path, as local NVMe is relative to a network download, that buffer can move to disk. It trades a scarce, contended resource (host memory shared on a GPU node with model weights and accelerator state) for an abundant, cheap one. The reassembly buffer in a container pull is one instance; log ingestion and shuffle stages in data-processing systems have the same shape, buffering in memory only to satisfy an ordering constraint. We have not explored these, but nothing about the mechanism is specific to container images: the technique involves pre-sizing files, direct-to-offset writing, and verification via re-reading, applicable to systems like Vector, Ray, and Apache Celeborn.

\subsection{The Next Bottleneck: Decompression}

Once download is parallel and the memory ceiling is gone, the cost hiding behind them shows up: decompression. Once a large layer arrives in seconds, decompression is the dominant remaining cost. We use \texttt{unpigz} for decompression, but it degrades to single-core inflation for a single large layer: container image layers are built as a single monolithic gzip stream rather than multiple independent stream members, so \texttt{unpigz} cannot parallelize within the layer. A 7~GiB layer that downloads in about five seconds therefore spends over a hundred seconds decompressing on one core. Multiple layers decompress in parallel, but a single large layer cannot be split across cores.

The disk-backed layout is what makes the next step possible. Because each layer is now materialized as a complete, seekable file, it can be handed to a parallel decompressor such as rapidgzip~\cite{knespel2023rapidgzip}, which builds an index over the compressed stream and inflates disjoint regions on separate cores.

\subsection{Applicability Limits}

Disk-backed assembly is not universally the right choice:
\begin{itemize}
    \item \emph{Small images.} For images below about a gigabyte, connection-setup overhead can outweigh the time saved by parallel fetch, and an ordinary sequential pull is adequate.
    \item \emph{Sparse-access workloads.} If a container touches only a sliver of its image, lazy loading skips the rest. DBPP targets the opposite case, the dense-access workload where most bytes are needed.
    \item \emph{Rotational or under-provisioned storage.} Writing chunks to arbitrary offsets assumes storage with no seek penalty. On NVMe or SSD, random-offset writes complete at the same throughput as sequential ones (roughly 3{,}000~MiB/s), so they stay off the critical path; on a spinning disk, forty writers hitting different offsets thrash the head, and disk-backed assembly can end up slower than buffering in memory.
\end{itemize}

\subsection{Limitations}
\begin{enumerate}
    \item \emph{Registry rate limiting.} Aggressive parallelism can trigger server-side rate limits; our system uses a static concurrency setting rather than adapting to registry feedback.
    \item \emph{Storage amplification.} During a pull, both the compressed layer and its decompressed contents are on disk at once, requiring transiently up to about twice the compressed image size.
    \item \emph{Serial per-layer decompression.} As above, a single layer still decompresses on one core; only cross-layer parallelism is exploited today.
    \item \emph{Single-registry evaluation.} All measurements use ECR; other registries may differ in rate-limiting and range-request behavior.
    \item \emph{Two-codebase comparison.} We compare containerd's in-memory transfer path against the SOCI snapshotter, two separate implementations. The controlled setup (Table~\ref{tab:controlled}) isolates the buffering strategy, but we cannot rule out every incidental implementation difference.
\end{enumerate}

%% file: sections/conclusion.tex
\section{Conclusion}
\label{sec:conclusion}

Container image pull does not need to hold the image in memory. DBPP writes each downloaded chunk straight to its offset on disk instead of piling chunks in the heap to reassemble them in order, so runtime memory stays flat regardless of image size, and every layer's digest still gets verified as the OCI format requires. Because the bytes already sit on disk as a complete, seekable file, DBPP verifies the compressed-layer SHA-256 and decompresses for the DiffID simultaneously: two passes containerd (in-memory) must run one after the other. Throughput stays comparable (within 12\%), and on memory-constrained nodes DBPP completes the pull where containerd is OOM-killed.

With download parallelized and memory bounded, the biggest cost now is single-core decompression of the large layers. That same on-disk layout is exactly what a parallel decompressor needs, so the decompression tail is the natural next target. Two smaller wins remain on the download side: more registry connections to close the throughput headroom, and concurrency that adapts to registry and storage feedback instead of being fixed in advance.

The pattern is not specific to container images. Any pipeline that buffers data in memory solely to satisfy an ordering or reassembly constraint can apply the same technique: pre-size the destination, write each piece directly to its offset, and verify by re-reading. The cost of the extra read stays off the critical path as long as the backing store is fast relative to the ingest rate, a condition that local NVMe meets comfortably against today's network speeds.

%% file: references.bib
@inproceedings{harter2016slacker,
  title={Slacker: Fast Distribution with Lazy Docker Containers},
  author={Harter, Tyler and Salmon, Brandon and Liu, Rose and Arpaci-Dusseau, Andrea C. and Arpaci-Dusseau, Remzi H.},
  booktitle={14th USENIX Conference on File and Storage Technologies (FAST '16)},
  pages={181--195},
  year={2016}
}

@inproceedings{brooker2023lambda,
  title={On-demand Container Loading in {AWS Lambda}},
  author={Brooker, Marc and Danilov, Mike and Greenwood, Chris and Piwonka, Phil},
  booktitle={2023 USENIX Annual Technical Conference (ATC '23)},
  year={2023}
}

@inproceedings{chen2022starlight,
  title={Starlight: Fast Container Provisioning on the Edge and over the {WAN}},
  author={Chen, Jun Lin and Liaqat, Daniyal and Gabel, Moshe and de Lara, Eyal},
  booktitle={19th USENIX Symposium on Networked Systems Design and Implementation (NSDI '22)},
  year={2022}
}

@inproceedings{faasnet2021,
  title={{FaaSNet}: Scalable and Fast Provisioning of Custom Serverless Container Runtimes at Alibaba Cloud Function Compute},
  author={Wang, Ao and Chang, Shuai and Tian, Huangshi and Wang, Hongqi and Yang, Haoran and Li, Huiba and Du, Rui and Cheng, Yue},
  booktitle={2021 USENIX Annual Technical Conference (ATC '21)},
  year={2021}
}

@inproceedings{li2020dadi,
  title={{DADI}: Block-Level Image Service for Agile and Elastic Application Deployment},
  author={Li, Huiba and Yuan, Yifan and Du, Rui and Ma, Kai and Liu, Lanzheng and Hsu, Windsor},
  booktitle={2020 USENIX Annual Technical Conference (ATC '20)},
  year={2020}
}

@inproceedings{flacio2025,
  title={{FlacIO}: Flat and Collective {I/O} for Container Image Service},
  author={Liu, Yubo and Li, Hongbo and Liu, Mingrui and Jing, Rui and Guo, Jian and Zhang, Bo and Guo, Hanjun and Ren, Yuxin and Jia, Ning},
  booktitle={23rd USENIX Conference on File and Storage Technologies (FAST '25)},
  pages={87--101},
  year={2025}
}

@inproceedings{knespel2023rapidgzip,
  title={Rapidgzip: Parallel Decompression and Seeking in Gzip Files Using Cache and Huffman Table},
  author={Knespel, Maximilian and Steinbach, Peter},
  booktitle={Proc. ACM HPDC},
  year={2023}
}

@misc{dragonfly2023,
  title={Dragonfly: An Intelligent {P2P}-Based Image and File Distribution System},
  author={{Dragonfly Authors}},
  year={2023},
  howpublished={[Online]. Available: \url{https://d7y.io}},
  note={Accessed: 2026-06-16}
}

@misc{kraken2019,
  title={Kraken: {P2P} Docker Registry},
  author={{Uber Engineering}},
  year={2019},
  howpublished={[Online]. Available: \url{https://github.com/uber/kraken}},
  note={Accessed: 2026-06-16}
}

@misc{soci2023,
  title={{SOCI} Snapshotter: Seekable {OCI} for Lazy Loading Container Images},
  author={{AWS Labs}},
  year={2023},
  howpublished={[Online]. Available: \url{https://github.com/awslabs/soci-snapshotter}},
  note={Accessed: 2026-06-16}
}

@misc{nydus2023,
  title={Nydus: Container Image Service for Accelerated Container Launch},
  author={{Nydus Authors}},
  year={2023},
  howpublished={[Online]. Available: \url{https://nydus.dev}},
  note={Accessed: 2026-06-16}
}

@misc{estargz2020,
  title={Startup Containers in Lightning Speed with Lazy Image Distribution on Containerd},
  author={{containerd/stargz-snapshotter Authors}},
  year={2020},
  howpublished={[Online]. Available: \url{https://github.com/containerd/stargz-snapshotter}},
  note={Accessed: 2026-06-16}
}

@misc{ociimage2023,
  title={{OCI} Image Format Specification, v1.1},
  author={{Open Container Initiative}},
  year={2023},
  howpublished={[Online]. Available: \url{https://github.com/opencontainers/image-spec}},
  note={Accessed: 2026-06-16}
}

@misc{ocidist2023,
  title={{OCI} Distribution Specification, v1.1},
  author={{Open Container Initiative}},
  year={2023},
  howpublished={[Online]. Available: \url{https://github.com/opencontainers/distribution-spec}},
  note={Accessed: 2026-06-16}
}

@misc{containerd2024,
  title={containerd: An Industry-Standard Container Runtime},
  author={{containerd Authors}},
  year={2024},
  howpublished={[Online]. Available: \url{https://github.com/containerd/containerd}},
  note={v2.2.4. Accessed: 2026-06-16}
}

@misc{moby2024,
  title={Moby: Docker Engine Container Runtime},
  author={{Moby Project}},
  year={2024},
  howpublished={[Online]. Available: \url{https://github.com/moby/moby}},
  note={Accessed: 2026-07-03}
}

@misc{containersimage2024,
  title={{containers/image}: Container Image Copy and Inspection Library ({skopeo}, {Podman})},
  author={{containers Project}},
  year={2024},
  howpublished={[Online]. Available: \url{https://github.com/containers/image}},
  note={Accessed: 2026-07-03}
}

@misc{rocm2024,
  title={{ROCm} {PyTorch} Training Container Images},
  author={{Advanced Micro Devices}},
  year={2024},
  howpublished={[Online]. Available: \url{https://hub.docker.com/r/rocm/pytorch}},
  note={Accessed: 2026-06-16}
}

@misc{awssocipull2025,
  title={Introducing {SOCI} Parallel Pull Mode for {Amazon EKS}},
  author={{Amazon Web Services}},
  year={2025},
  howpublished={[Online]. Available: \url{https://aws.amazon.com/blogs/containers/introducing-seekable-oci-parallel-pull-mode-for-amazon-eks/}},
  note={Accessed: 2026-06-16}
}

@misc{eksautomode2025,
  title={{re:Invent} 2025: {Amazon EKS Auto Mode} --- Evolving {Kubernetes} Operations},
  author={{Amazon Web Services}},
  year={2025},
  howpublished={[Online]. Available: \url{https://repost.aws/articles/ARH8G-CCrCTram8sa017ezPw/}},
  note={Accessed: 2026-06-16}
}

@misc{eksultrascale2025,
  title={Under the Hood: {Amazon EKS} Ultra Scale Clusters},
  author={{Amazon Web Services}},
  year={2025},
  howpublished={[Online]. Available: \url{https://aws.amazon.com/blogs/containers/under-the-hood-amazon-eks-ultra-scale-clusters/}},
  note={Accessed: 2026-06-16}
}

@misc{appsflyer2025soci,
  title={How We Cut {GitHub Actions} Runner Cold Start by 82\% on {EKS} with {SOCI} Parallel Pull},
  author={{AppsFlyer Engineering}},
  year={2025},
  howpublished={[Online]. Available: \url{https://medium.com/appsflyerengineering/how-we-cut-github-actions-runner-cold-start-by-82-on-eks-with-soci-parallel-pull-fe8a44faf313}},
  note={Accessed: 2026-06-16}
}

@misc{grab2025lazyloading,
  title={Docker Lazy Loading at {Grab}},
  author={{Grab Engineering}},
  year={2025},
  howpublished={[Online]. Available: \url{https://engineering.grab.com/docker-lazy-loading}},
  note={Accessed: 2026-06-16}
}

@misc{aws2024bottlerocket,
  title={Reduce Container Startup Time on {Amazon EKS} with {Bottlerocket} Data Volume},
  author={{Amazon Web Services}},
  year={2024},
  howpublished={[Online]. Available: \url{https://aws.amazon.com/blogs/containers/reduce-container-startup-time-on-amazon-eks-with-bottlerocket-data-volume/}},
  note={Accessed: 2026-06-16}
}
